\documentclass[a4paper,11pt]{article}
\pdfoutput=1 

\usepackage{jheppub} 

\usepackage[T1]{fontenc} 
\usepackage{comment}
\usepackage{makecell}

\usepackage{graphicx}
\usepackage{bm}
\usepackage{color}
\usepackage{ulem}
\usepackage{siunitx}

\usepackage {cancel}

\usepackage{orcidlink}

\usepackage{threeparttable}

\newcommand{\MMS}{M_{\rm recoil}^2}
\newcommand{\mrec}{M_{\rm recoil}}
\newcommand{\yones}{\Upsilon(1S)}
\newcommand{\ytwos}{\Upsilon(2S)}

\newcommand{\yfos}{\Upsilon(4S)}

\newcommand{\yns}{\Upsilon(nS)}
\newcommand{\yonetwos}{\Upsilon(1,2S)}

\newcommand{\eff}{\varepsilon}
\newcommand{\BR}{{\cal B}}

\newcommand{\pip}{\pi^+}
\newcommand{\pim}{\pi^-}
\newcommand{\piz}{\pi^0}

\newcommand{\kl}{K_L^0}

\newcommand{\jpsi}{J/\psi}

\newcommand{\psp}{\psi(2S)}

\newcommand{\pbar}{\bar{p}}
\newcommand{\pc}{P_c^{+}}

\newcommand{\pca}{P_c(4312)^+}
\newcommand{\pcb}{P_c(4440)^+}
\newcommand{\pcc}{P_c(4457)^+}
\newcommand{\pjpsi}{pJ/\psi}
\newcommand{\pbjpsi}{\bar{p}J/\psi}

\newcommand{\EE}{e^+e^-}
\newcommand{\MM}{\mu^+\mu^-}
\newcommand{\LL}{\ell^+\ell^-}

\newcommand{\qqb}{q\bar{q}}

\newcommand{\ppjpsi}{\pi^+\pi^- J/\psi}

\newcommand{\bcntr}{\begin{center}}
\newcommand{\ecntr}{\end{center}}
\newcommand{\beq}{\begin{equation}}
\newcommand{\eeq}{\end{equation}}
\newcommand{\beqar}{\begin{eqnarray}}
\newcommand{\eeqar}{\end{eqnarray}}
\newcommand{\bitm}{\begin{itemize}}
\newcommand{\benu}{\begin{enumerate}}

\newcommand{\bitmb}{\begin{itemize}}
\newcommand{\benub}{\begin{enumerate}}
\newcommand{\eitm}{\end{itemize}}

\newcommand{\bfrm}{\begin{frame}}
\newcommand{\efrm}{\end{frame}}
\newcommand{\bct}{\begin{center}}
\newcommand{\ect}{\end{center}}
\newcommand{\bclm}{\begin{columns}}
\newcommand{\eclm}{\end{columns}}
\newcommand{\bpic}{\begin{overpic}}
\newcommand{\epic}{\end{overpic}}
\newcommand{\bblk}{\begin{block}}
\newcommand{\eblk}{\end{block}}

\newcommand{\eenu}{\end{enumerate}}

\newcommand{\cm}{\si{\centi\metre}}

\newcommand{\fb}{\si{\femto\barn}}

\newcommand{\infb}{\fb^{-1}}

\newcommand{\gevcs}{\hbox{GeV}/c^2}
\newcommand{\gevcss}{\hbox{GeV}^2/c^4}
\newcommand{\gevc}{\hbox{GeV}/c}
\newcommand{\gev}{\hbox{GeV}}
\newcommand{\mevcs}{\hbox{MeV}/c^2}

\newcommand{\mev}{\hbox{MeV}}

\newcommand{\reduline}{\bgroup\markoverwith
{\textcolor{red}{\rule[0.5ex]{2pt}{0.4pt}}}\ULon}

\def\Journal#1#2#3#4{{#1} {\bf #2}, #3 (#4)}

\def\NIMA{Nucl. Instrum. Methods A}
\def\PL{Phys. Lett.}
\def\PLB{Phys. Lett. B}
\def\PRL{Phys. Rev. Lett.}
\def\PRD{Phys. Rev. D}

\def\CPC{Chin. Phys. C}
\def\EPJC{Eur. Phys. J. C}
\def\PTEP{Prog. Theor. Exp. Phys. }
\def\ARNPS{Annu. Rev. Nucl. Part. Sci.}

\def\JHEP{J. High Energy Phys. }

\def\Mod{Rev. Mod. Phys.}

\preprint{} \preprint{ \vbox{ \hbox{   }
		\hbox{Belle Preprint 		2024-02	}
		\hbox{KEK Preprint 		2023-54	}
}}

\title{
\quad\\[1.0cm]
Search for a pentaquark state decaying into $\pjpsi$ in $\yonetwos$ inclusive decays at Belle}

  \author{X.~Dong\,\orcidlink{0000-0001-8574-9624}} 
  \author{S.~M.~Zou\,\orcidlink{0000-0003-3377-7222}}
  \author{H.~Y.~Zhang\,\orcidlink{0000-0001-6918-4029}} 
  \author{X.~L.~Wang\,\orcidlink{0000-0001-5805-1255}} 
  \author{I.~Adachi\,\orcidlink{0000-0003-2287-0173}} 
  \author{J.~K.~Ahn\,\orcidlink{0000-0002-5795-2243}} 
  \author{H.~Aihara\,\orcidlink{0000-0002-1907-5964}} 
  \author{S.~Al~Said\,\orcidlink{0000-0002-4895-3869}} 
  \author{D.~M.~Asner\,\orcidlink{0000-0002-1586-5790}} 
  \author{H.~Atmacan\,\orcidlink{0000-0003-2435-501X}} 
% ...
\collaboration{The Belle Collaboration}

\date{\today}

\abstract{
Using the data samples of 102 million $\yones$ and 158 million $\ytwos$ events collected by the Belle detector, we 
search for a pentaquark state in the $\pjpsi$ final state from $\yonetwos$ inclusive decays. Here, the 
charge-conjugate $\pbjpsi$ is included. We observe clear $\pjpsi$ production in $\yonetwos$ decays and measure the 
branching fractions to be $\BR[\yones\to \pjpsi + anything] = [8.1 \pm 0.6(stat.) \pm 0.5(syst.)] \times
10^{-5}$ and $\BR[\ytwos \to \pjpsi + anything] = [4.3 \pm 0.5(stat.) \pm 0.4(syst.)] \times 10^{-5}$. We also
measure the cross section of inclusive $\pjpsi$ production in $\EE$ annihilation to be $\sigma(\EE \to \pjpsi
+ anything) = [108\pm 11 (stat.) \pm 6(syst.)]~\fb$ at $\sqrt{s} = 10.52~\gev$ using an $89.5~\infb$ continuum
data sample. There is no significant $\pca$, $\pcb$ or $\pcc$ signal found in the $\pjpsi$ final states in 
$\yonetwos$ inclusive decays. We determine the upper limits of $\BR[\yonetwos\to \pc + anything] \cdot \BR(\pc\to 
\pjpsi)$ to be at the $10^{-6}$ level.}

\keywords{$\EE$ Experiments, Bottomonium, Exotic State, Pentaquark State}

\begin{document}

\maketitle
\flushbottom

\section{Introduction}

In the conventional quark model, a hadron is either a meson containing a quark and an anti-quark or an (anti-)baryon 
containing three (anti-)quarks. However, the fundamental theory of strong interaction, Quantum Chromodynamics, does 
not forbid new structures of hadrons beyond the conventional quark model, such as glueball states containing only 
gluons,  hybrid states containing gluons and quarks, or multi-quark states containing more than three 
quarks~\cite{quark-model}. Many theoretical and experimental efforts have been devoted to predicting and searching 
for these exotic states~\cite{epjc, Nonstandard}. In 2003, the Belle experiment observed the $X(3872)$ in $B \to K + 
\ppjpsi$ decay~\cite{x3872}, which was the earliest evidence yet of the existence of exotic states. Five years
later, when studying the decay $B\to K + \pip \psp$, Belle observed the $Z(4430)^+$~\cite{z4430}, which is
electrically charged and evidence for a four-quark meson~\cite{Nonstandard}.  Since then, many candidate multi-quark
states have been observed by the Belle, LHCb, and BESIII experiments~\cite{zc-lhcb, z1z2, zb, zc-bes, zc-belle, 
zc4020, zc_4050_belle, zc_4050_bes3}. In the pentaquark sector, the LHCb experiment discovered $P_c(4380)^+$ and 
$P_c(4450)^+$ in the decay $\Lambda_b \to K + \pjpsi$~\cite{pc}, but an updated analysis using ten times the 
statistics divided the structures into three states~\cite{lhcb_pc_2019}, the $\pca$, $\pcb$, and $\pcc$. The
deuteron can be considered a candidate for a hexaquark state~\cite{deuteron}. The observations of deuterons in the
$\yns$ inclusive decays by the ARGUS, CLEO, and BaBar experiments provide clues for searching for more candidates of
multi-quark states in the $\yns$ inclusive decays~\cite{cleo_db_yns, ARGUS_db_yns, Babar_db_yns}.

The Belle experiment collected the world's largest $\yonetwos$ data samples in  its last operation years before the
KEKB accelerator was shut down in 2010. The $\yones$ data sample with an integrated luminosity $\mathcal{L}_{\yones}
= 5.8~\infb$ contains $(102\pm 2) \times 10^6~\yones$ events~\cite{y1sevnt}, while the $\ytwos$ data sample has
$\mathcal{L}_{\ytwos} = 24.7~\infb$ and $(158 \pm 4)\times 10^6~\ytwos$ events~\cite{y2sevnt}. Using the two data
samples, we search for a $\pc$ state in the inclusive production of $\pjpsi$ final states via $\yonetwos$ decays.
Here and hereinafter, $\pc$ is $\pca$, $\pcb$, or $\pcc$. The charge-conjugated final state $P_c^- \to \bar{p}
\jpsi$ is included throughout this study. We also use a Belle continuum data sample with an integrated luminosity
of $\mathcal{L}_{\rm cont} = 89.5~\infb$ taken at center-of-mass (c.m.) energy $\sqrt{s} = 10.52~\gev$ [$60~\mev$
below the peak of the $\yfos$ resonance] to investigate the $\pjpsi$ final state from continuum productions, which
could be backgrounds in the $\yonetwos$ data samples for studying the $\yonetwos$ decays.

\section{The Belle Detector and Monte Carlo simulation}

The Belle detector is a large-solid-angle magnetic spectrometer~\cite{Belle}. It consists of several subdetectors, 
including a silicon vertex detector, a central drift chamber with 50 layers, an array of aerogel threshold Cherenkov 
counters, a barrel-like arrangement of time-of-flight scintillation counters, and an electromagnetic calorimeter 
(ECL) comprised of CsI(Tl) crystals. All the above are located within a superconducting solenoid coil which
generates a magnetic field of 1.5 T. An iron flux return outside the coil is instrumented to detect $\kl$ mesons
and identify muons. The origin of the coordinate system is defined as the position of the nominal interaction point.
The $z$ axis is aligned with the direction opposite to the $e^{+}$ beam and points along the magnetic field within
the solenoid. The $x$ axis points horizontally outwards of the storage ring, and the $y$ axis is vertically
upwards. The angles of the polar ($\theta$) and azimuthal ($\phi$) are measured relative to the positive $z$ and
$x$ axes.

To optimize the selection criteria, we use EvtGen to simulate signal Monte Carlo (MC) samples of $\yonetwos \to \pc + 
\bar{n}/\pbar + q\bar{q}^\prime$ with $\pc\to \pjpsi$ according to three-body phase space~\cite{evtgen}, where $q\bar{q}^\prime~(q, q^\prime =u,d,s,c)$ is a 
quark-antiquark pair of random flavor whose hadronization is simulated by PYTHIA6.4~\cite{pythia}. 
Each $\pc$ MC sample has $2\times 10^4$ events, and we combine the three $\pc$ signal MC samples for the selection criteria 
optimization. To study the efficiency and mass resolution of the $\pjpsi$ invariant mass ($M_{\pjpsi}$), we generate 
efficiency MC samples of $\pc$, whose mass is fixed to different values from $4.1~\gevcs$ to $5.0~\gevcs$, and the 
width is set to zero. To study $\pjpsi$ production not due to $\pc$ decays, we generate a no-$\pc$ MC sample of 
$\yonetwos\to \jpsi + p +\pbar/\bar{n} + \qqb$ according to four-body phase space~\cite{evtgen}. 
In all these $\pc$ MC samples and no-$\pc$ MC samples, the relative fractions 
of the $\jpsi p\bar{p}$ and $\jpsi p\bar{n}$ channels are determined according to their multiple combination ratios extracted from 
data, with measured values of 7.8\% for $\yonetwos$ and 8.3\% for continuum. Typically, each $\jpsi p\bar{p}$ event has two combinations, and the  $\jpsi p\bar{n}$ events have rare multiple combinations. To simulate the 
hadronization of $q\bar{q}^\prime$, we define a state of $X \to q\bar{q}^\prime$ where $X$ has a mass of $2.6~\gevcs$ and a width of 
$2.7~\gev$ in $\yones$ decays; similarly a mass of $2.4~\gevcs$ and width of $3.3~\gev$ in $\ytwos$ decays. We
simulate the geometry and the response of the Belle detector using a GEANT3-based MC technique~\cite{geant3}.

\section{Event selection }

To reconstruct the $\pjpsi$ final state, we select events with at least three well-measured charged tracks. Two 
tracks with opposite charges are chosen as candidates for $\jpsi$ decaying into $\EE$ (called the $\EE$ mode) or 
$\MM$ (called the $\MM$ mode). A well-measured charged track has impact parameters of $dr < 0.5~\cm$ in the $r-\phi$ 
plane and $|dz| < 5~\cm$ in the $r-z$ plane with respect to the interaction point, and a transverse momentum larger 
than $0.1~\gevc$. For each charged track, we combine information from the subdetectors of Belle to form a likelihood 
$\mathcal{L}_i$ for each putative particle species ($i$)~\cite{pid}. We form the likelihood ratios $\mathcal{R}_e 
\equiv \mathcal{L}_e/(\mathcal{L}_e + \mathcal{L}_{\rm hadrons})$ and $\mathcal{R}_\mu \equiv 
\mathcal{L}_\mu/(\mathcal{L}_\mu+\mathcal{L}_{\rm hadrons})$ for electron and muon identifications~\cite{EID, MUID}. 
For electrons from $\jpsi\to \EE$ decay, we require both tracks to have $\mathcal{R}_e > 0.9$ and include the 
bremsstrahlung photons detected in the ECL within 0.05 radians of the original $e^+$ or $e^-$ direction in 
calculating the $\EE(\gamma)$ invariant mass. For muons from $\jpsi\to \MM$ decay, we require both tracks to have 
$\mathcal{R}_\mu > 0.9$. The single lepton identification efficiency is $(93.9 \pm 0.2)\%$ in the $\EE$ mode and 
$(91.9 \pm 0.2)\%$ in the $\MM$ mode. We identify a track with $\mathcal{R}_{p/K} = \frac{\mathcal{L}_p} 
{\mathcal{L}_p + \mathcal{L}_K} > 0.6$ and $\mathcal{R}_{p/\pi} = \frac{\mathcal{L}_p} {\mathcal{L}_p + 
\mathcal{L}_\pi} > 0.6$ as a proton. To remove the proton
candidates from beam backgrounds, we require the difference of the $dz$ parameter for $p$ and $\ell^{\pm}$ to be
$|\Delta dz| < 0.5~\cm$. The efficiency of proton identification is $(97.3 \pm 0.1)\%$.

We study the backgrounds of the proton from a secondary hadron's decay. Final states of many baryons, such as
$\Sigma^0$, $\Xi$, $\Omega$ and excited $\Lambda$, contain a $\Lambda$. To remove backgrounds from $\Lambda \to
p\pi$ decay in the proton selection, we reconstruct all the pion candidates with $\mathcal{R}_{\pi/K} =
\mathcal{L}_\pi/(\mathcal{L}_\pi + \mathcal{L}_K) > 0.6$ and a charge opposite to that of the  proton. The number of
candidates is about 40, so that the background level is low. We remove the proton candidate if it is part of any
$p\pi$ combination of mass $1.105~\gevcs < M_{p\pi} < 1.12~\gevcs$, where $M_{p\pi}$ is the invariant mass of the
$p\pi$ combination. According to isospin symmetry in strong interaction, we expect the backgrounds containing
$\Sigma \to p\piz$ can be ignored too. A $\Delta$ has a large width and decays at the interaction point. We see
no obvious $\Delta$ signal in the $p\pim$ invariant mass distributions. We conclude that the background level due to
a proton from a secondary particle decay is very low in estimating the production of $p\jpsi$ in $\yonetwos$
inclusive decays or the $\EE$ continuum production, and they do not contribute peaking backgrounds in the $p\jpsi$
invariant mass distributions. This background has a negligible effect in estimating the $\pc$ productions.

\begin{figure}[tbp]
\centering
\psfig{file=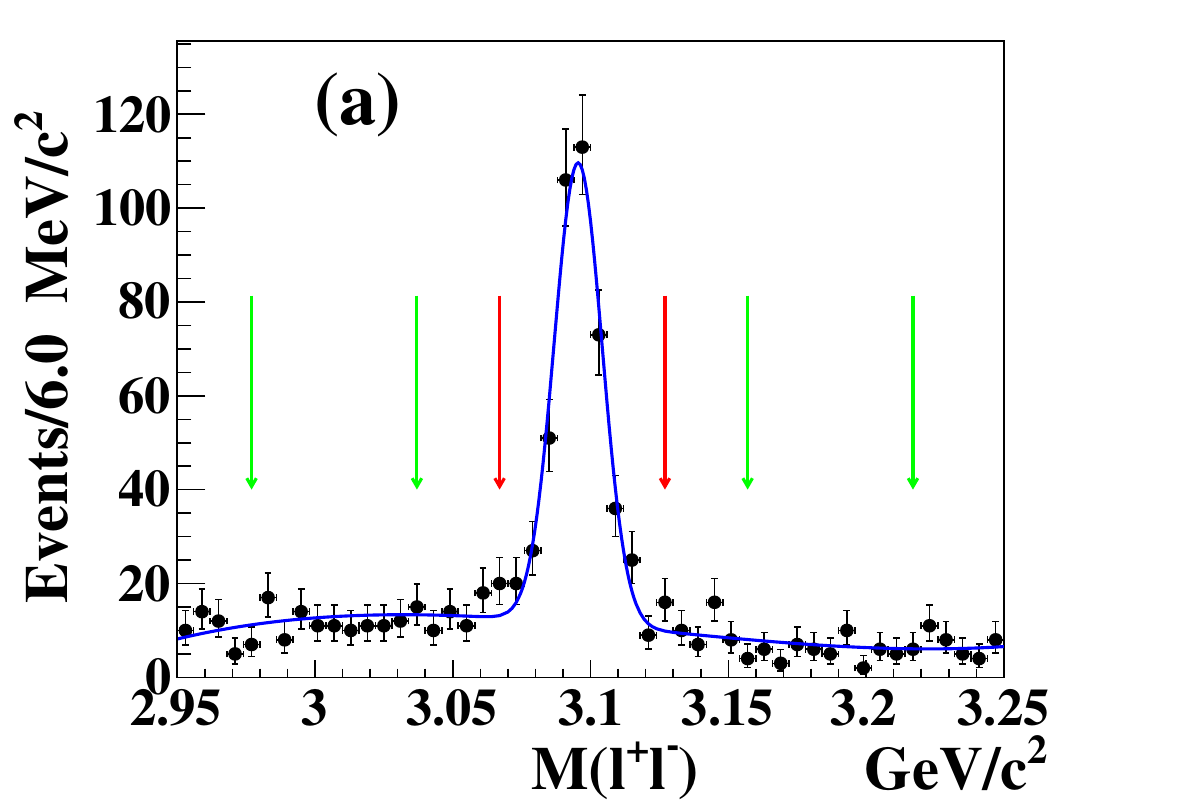,width=.45\textwidth}
\psfig{file=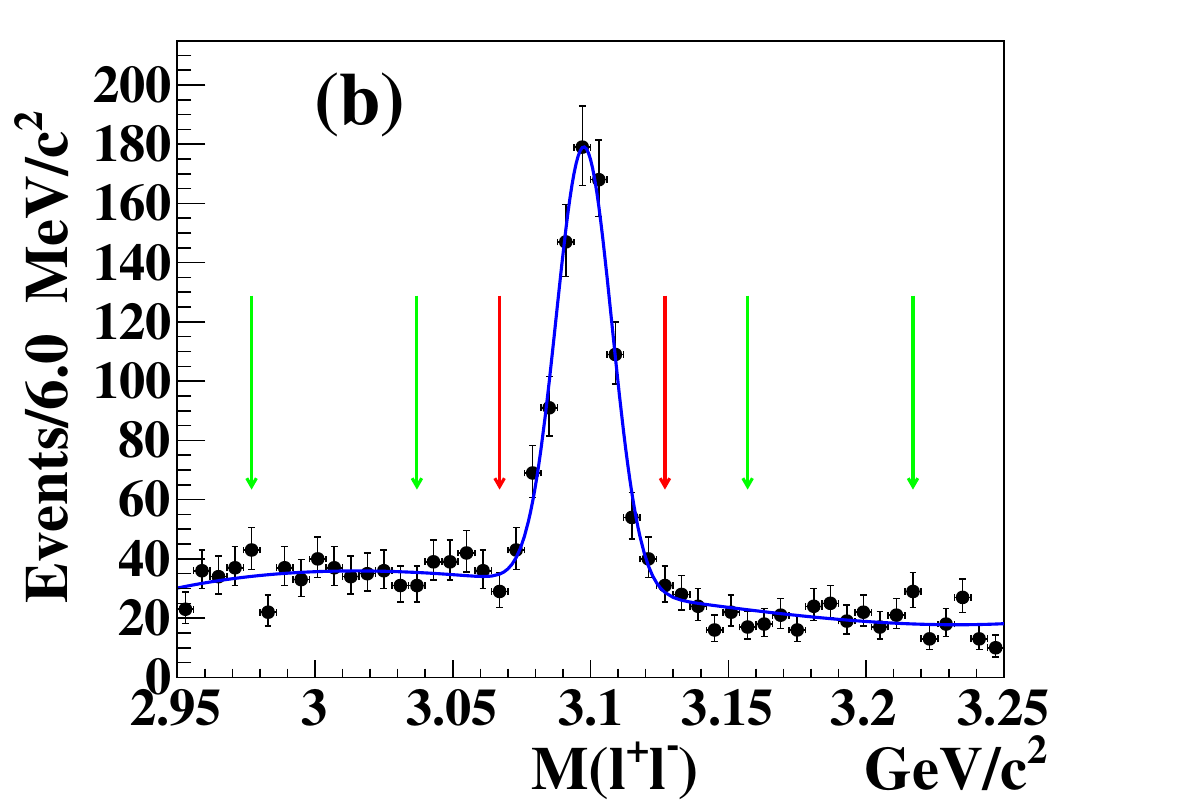,width=.45\textwidth}
\caption{The invariant-mass distributions of the lepton pair from (a) the $\yones$ data sample and (b) the $\ytwos$ 
data sample. The curves show the best fit results with a Gaussian function for the $\jpsi$ signal and a second-order 
Chebyshev function for the backgrounds. The red arrows indicate the $\jpsi$ signal region and the green ones 
indicate the $\jpsi$ mass sideband regions.}
\label{mll}
\end{figure}

The $\yonetwos$ data samples and the continuum data sample show clear $\jpsi$ signals in the $\EE$ mode 
and the $\MM$ mode. Figure~\ref{mll} shows the invariant-mass distributions of the lepton pair ($M_{\LL}$), which is 
the sum of the $\EE$ mode and the $\MM$ mode, in the $\yonetwos$ data samples. Fitting the $M_{\LL}$ distributions 
using a Gaussian function for the $\jpsi$ signal and a second-order Chebychev function for the backgrounds, we get 
the mass resolution of the $\jpsi$ signal to be $8.7 \pm 0.6~\mevcs$ ($10.1 \pm 0.5~\mevcs$) in the $\yones$
[$\ytwos$] data sample and $8.9 \pm 0.2~\mevcs$ ($10.0 \pm 0.2~\mevcs$) in the signal MC simulation of $\yones$
[$\ytwos$] decays. We define the $\jpsi$ signal region to be $|M_{\LL} - m_{\jpsi}| < 3\sigma$, where $m_{\jpsi}$ is 
the nominal mass of $\jpsi$~\cite{PDG} and $\sigma = 10~\mevcs$. To estimate the backgrounds in the $\jpsi$
selection, we define the $\jpsi$ mass sideband regions as $|M_{\LL} - m_{\jpsi} + 9\sigma| < 3\sigma$ and $|M_{\LL}
- m_{\jpsi} - 9\sigma| < 3\sigma$.

\begin{figure}[tbp]
\centering
\psfig{file=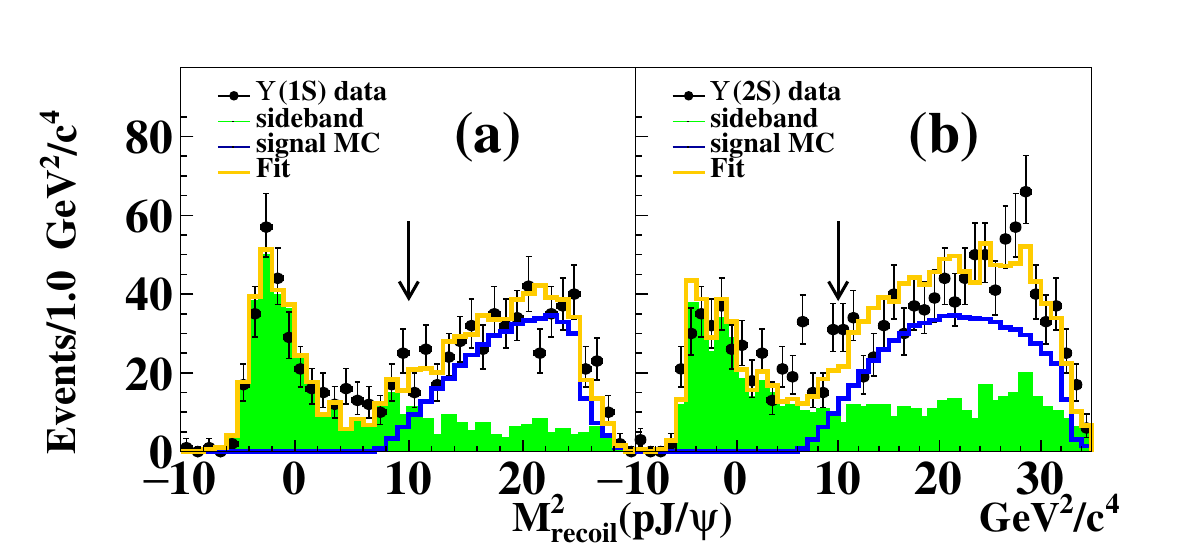,width=.8\textwidth}
\psfig{file=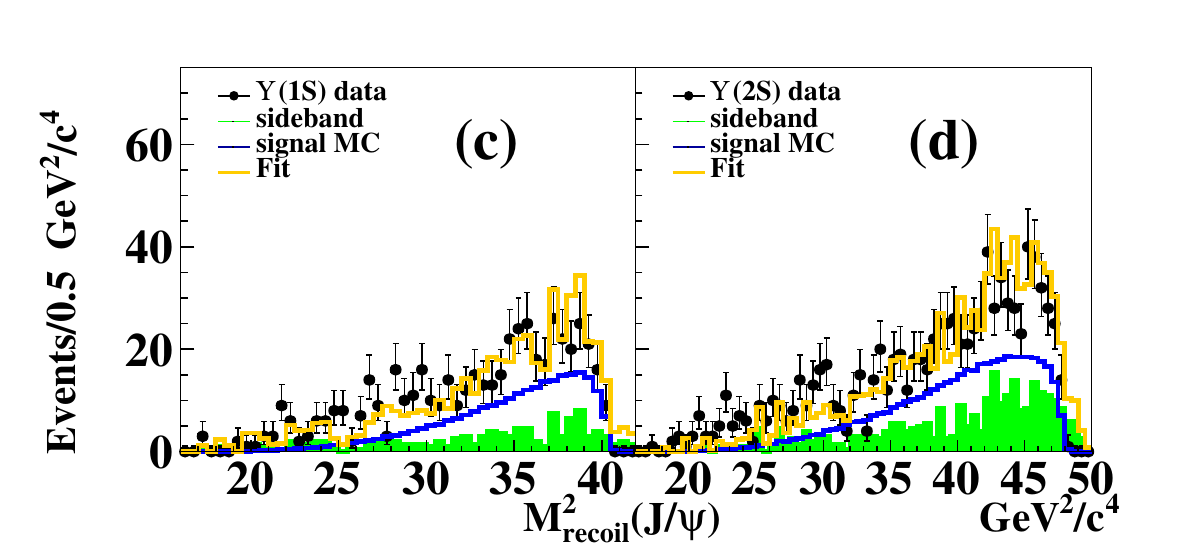, width=.8\textwidth}
\caption{The distributions of the recoil mass squared of $\pjpsi$ (upper),  and $\jpsi$ (lower) in $\yones$ (left)
and $\ytwos$ (right) decays. The dots with error bars are data, the shaded histograms are backgrounds estimated from 
the $\jpsi$ mass sideband regions, and the solid blue histograms are signal MC simulations. The yellow histograms
show the fit results with the signal MC simulations and fixed backgrounds estimated from the $\jpsi$ mass sidebands.
The arrows show the requirement $\MMS(\pjpsi) > 10~{\gevcss}$, which  has been applied for the $\MMS(\jpsi)$ distributions.}
\label{mrec} 
\end{figure}

Figures~\ref{mrec}(a) and \ref{mrec}(b) show the distributions of the recoil mass squared against the $\pjpsi$
system in $\yonetwos$ data samples and signal MC simulations. This quantity is calculated by $\MMS(\pjpsi) \equiv
(P_{\EE} - P_{\pjpsi})^2$, where $P_{\EE}$ is the 4-momentum of the $\EE$ collision and $P_{\pjpsi}$ is the
4-momentum of the $\pjpsi$ combination. In data, there are accumulations between $-5~\gevcss$ and $5~\gevcss$ for
the events selected in the $\jpsi$ signal region, and these can be described well with the backgrounds estimated
from the $\jpsi$ mass sideband regions. These backgrounds appear in the $\EE$ mode but are scarce in the $\MM$
mode. On the other hand, these events produce a large peak at zero and a wide distribution of the recoil mass
squared against the $\jpsi$ candidate, calculated by $\MMS(\jpsi) \equiv (P_{\EE} - P_{\jpsi})^2$, where
$P_{\jpsi}$ is the 4-momentum of the $\jpsi$ candidate and used to calculate the $M_{\LL}$. They are
identified as backgrounds from Bhabha events with high energy bremsstrahlung radiation photon(s) and an additional
proton from beam backgrounds. As this proton is not from an $\EE$ collision, this background can produce negative
accumulations in the $\MMS(\pjpsi)$ distributions. We require $\MMS(\pjpsi) > 10~\gevcss$ to suppress these
backgrounds with a selection efficiency of about 99\% in $\yonetwos$ decays. Figures~\ref{mrec}(c) and
\ref{mrec}(d) show the distributions of $\MMS(\jpsi)$ after this requirement. We fit the data with the
histograms obtained from the signal MC simulations and fixed backgrounds estimated from the $\jpsi$ mass sidebands.
As shown in Fig.~\ref{mrec}, the fits yield good agreements.

\section{Invariant mass spectra of $\pjpsi$}

\begin{figure}[tbp]
\centering
\psfig{file=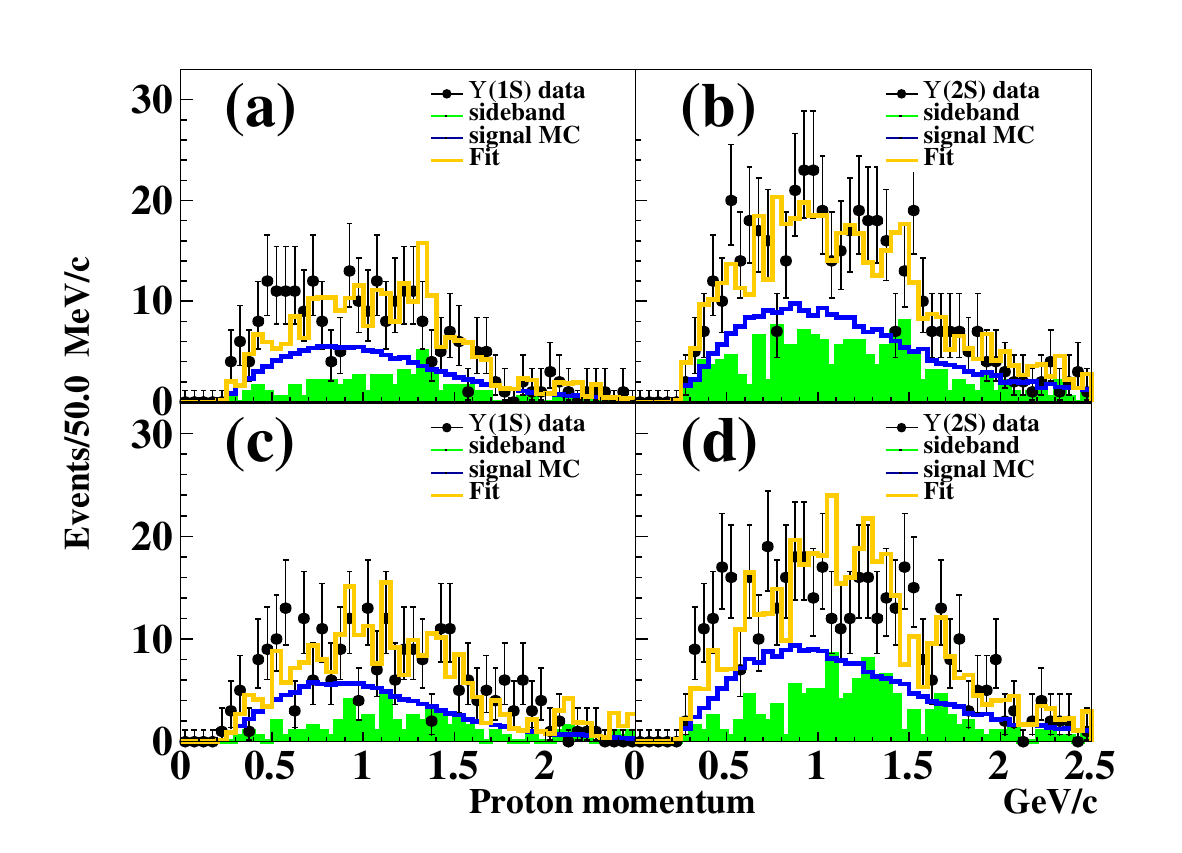,width=.8\textwidth}
\caption{The momentum distributions of $p/\pbar$ in $\yonetwos$ inclusive decays. The first row is the momenta of 
$p$ and the second of $\pbar$. The left and right panels are $\yones$ and $\ytwos$, respectively. The dots with 
error bars are data, the shaded histograms are backgrounds estimated from the $\jpsi$ mass sideband regions, and
the solid histograms are signal MC simulations. The yellow histograms show the fit results with the signal MC
simulations and fixed backgrounds estimated from the $\jpsi$ mass sidebands. }
\label{ppbar}
\end{figure}

All the candidates satisfying the selection criteria described above are accepted, including $p$ or $\pbar$ with
the same $\jpsi$ candidate or multiple candidates sharing one lepton. We show the momentum distributions of the
$p/\pbar$ after selection criteria in Fig.~\ref{ppbar}.

According to the efficiency MC simulations, we obtain an efficiency varying from 29\% (28\%) to 36\% (34\%) in the
$\yones$ [$\ytwos$] decays, and the mass resolution increasing from $1.6~\mevcs$ to $4.9~\mevcs$ for $M_{\pjpsi} \in
[4.1,~5.0]~\gevcs$. Here, to avoid the broadening due to the mass resolution of the $\LL$ combination for a
$\jpsi$ candidate, we use the calculation $M_{\pjpsi} = M_{p\LL} - M_{\LL} + m_{\jpsi}$, where $M_{p\LL}$ is the
invariant masses calculated from $P_{\pjpsi}$. We notice that the width of $\pcc$ reported by LHCb is
$\Gamma_{\pcc} = 6.4 \pm 2.0^{+5.7}_{-1.9}~\mev$~\cite{lhcb_pc_2019} and the mass resolution near the mass of
$\pcc$ is about $3.0~\mevcs$. Therefore, we need to consider the mass resolution in fitting the $M_{\pjpsi}$
distributions for the possible $\pc$ signals. Here and hereinafter, the first uncertainty quoted is statistical,
while the second corresponds to the total systematic uncertainty.

We then study the $M_{\pjpsi}$ distributions from the signal MC simulations of $\pca$, $\pcb$, and $\pcc$. In each 
distribution, there is a clear $\pc$ peak and a plateau of wrong combination with particle(s) from the recoil of 
$\pc$. We perform a fit to this distribution using a Breit-Wigner function convolved with a Gaussian resolution 
function to describe the signals and a first-order polynomial function to describe the plateau of the wrong 
combinations. The fit range is $M_{\pc} \pm 200~\mevcs$, where $M_{\pc}$ is the mass of $\pc$. The fits yield mass 
resolutions of around $3~\mevcs$ for each $\pc$ state. The mass resolutions obtained here agree with those 
obtained from the efficiency MC simulations. We calculate the ratio $\mathcal{R} \equiv N_{\pc}/N_{\pjpsi}$
to be approximately 0.95, where the $N_{\pc}$ and $N_{\pjpsi}$ are the number of $\pc$ signals from the fit and the
number of all $\pjpsi$ combinations being selected between $4.0~\gevcs$ and $5.0~\gevcs$, respectively. The 
efficiencies of all combinations ($\eff^{\rm MC}_{\rm allcmb}$) are about 40\%. We list the details of the mass
resolutions, the ratios $\mathcal{R}$, and the efficiencies $\eff^{\rm MC}_{\rm allcmb}$ from the signal MC 
simulations of $\pca$, $\pcb$, and $\pcc$ in $\yonetwos$ inclusive decays in Table~\ref{detail_Pc_simulation}. 

\begin{table}
\begin{center}
\caption{The mass resolution, the ratio of the number of $\pc$ signals to the number of all $\pjpsi$ combinations, 
and the efficiency of all the $\pjpsi$ combinations from the signal MC simulations of $\pca$, $\pcb$, and $\pcc$ in 
$\yonetwos$ decays.  }
\begin{small}
\begin{tabular}{c | c | c | c | c | c | c }
\hline\hline 
        					&\multicolumn{3}{c|}{$\yones$ decays} &  \multicolumn{3}{c}{$\ytwos$ decays} \\\hline
   --- 						& $\pca$  & $\pcb$ & $\pcc$ & $\pca$ & $\pcb$ & $\pcc$   \\\hline
mass resolution ($\mevcs$)  &  $2.9\pm 0.1$   & $3.2\pm 0.2$  & $3.4\pm 0.1$  & $3.0\pm 0.1$
&  $3.2\pm 0.2$ & $3.4\pm 0.1$       \\
Ratio of $N_{\pc}/N_{\pjpsi}$    &  0.95   & 0.94   & 0.95  & 0.96  & 0.95  & 0.95       \\
$\eff^{\rm MC}_{\rm allcmb}$  (\%)  &  $39.7 \pm 0.1$   & $39.5\pm 0.1$  & $39.2 \pm 0.1$
&  $39.8\pm 0.1$  & $39.8\pm 0.1$ & $39.5 \pm 0.1$       \\\hline\hline
\end{tabular}
\end{small}
\label{detail_Pc_simulation}
\end{center}
\end{table}

\begin{figure}[tbp]
\centering
\includegraphics[width=1.0\textwidth]{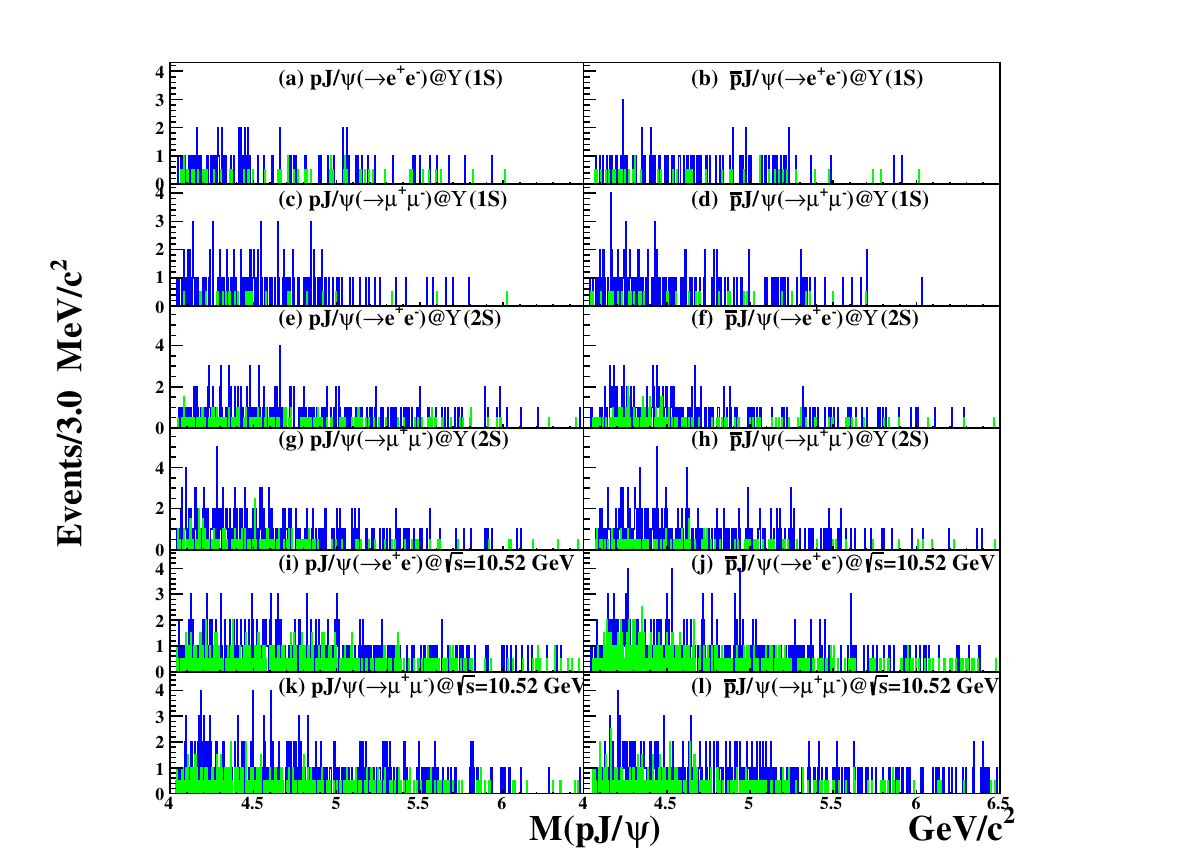}
\caption{The invariant-mass distributions of $\pjpsi$ in the $\yones$, $\ytwos$, and continuum data samples. The
solid histograms are the $\pjpsi$ signals, and the shaded histograms are backgrounds estimated from the $\jpsi$
mass sideband regions. }
\label{mpjpsi-bin2}
\end{figure}

We study the $M_{\pjpsi}$ distributions obtained from the $\yones$, $\ytwos$, and continuum data samples, and show 
them in Figs.~\ref{mpjpsi-bin2}(a-d), \ref{mpjpsi-bin2}(e-h), and \ref{mpjpsi-bin2}(i-l), respectively. There are 
clear $\pjpsi$ signals in the three data samples. As mentioned, we use the distributions obtained from the continuum 
data sample to estimate the backgrounds from $\EE$ annihilation in the $\yonetwos$ decays. For this, we scale the 
luminosities and correct for the efficiencies and the c.m. energy dependence of the Quantum Electrodynamics (QED) 
cross section $\sigma_{\EE} \propto 1/s$, resulting in scale factors $f_{\rm scale} = (\mathcal{L}_{\yonetwos}
\times \eff_{\yonetwos} \times s_{\rm cont})/(\mathcal{L}_{\rm cont} \times \eff_{\rm cont}\times s_{\yonetwos}) =
0.077$ and 0.303 for $\yones$ and $\ytwos$, respectively. We find no peaking component in the combined $M_{\pjpsi}$
distribution from Figs~\ref{mpjpsi-bin2}(i-l).  We obtain the number of $\pjpsi$ candidates to be
$N^{\pjpsi}_{\rm cont} = 397 \pm 34$ after subtracting the backgrounds estimated from the $\jpsi$ mass sideband
regions. To estimate the backgrounds due to the mis-identification of proton, we replace the proton identification
requirements with $\mathcal{L}_p/(\mathcal{L}_p+\mathcal{L}_K)< 0.4$ or $\mathcal{L}_p/(\mathcal{L}_p +
\mathcal{L}_\pi) < 0.4$ in the signal selection. We obtain $1746\pm 82$ $K^\pm\jpsi$ signals with kaon
identification efficiency of 93.5\% or $1710\pm 82$ $\pi^\pm\jpsi$ signals with pion identification efficiency of
92.4\%. Taking into account mis-identification rates, we expect the number of backgrounds from $K^\pm\jpsi$ or
$\pi^\pm\jpsi$ to be $50.3\pm 2.4 \pm 2.4$, where the systematic uncertainty is described in Sec.~\ref{syserr}.
Hence, the number of $\pjpsi$ events after all background subtractions is found to be $N^{\pjpsi}_{\rm cont} = 347
\pm 34$. With the scale factor $f_{\rm scale}$, we expect  $27 \pm 3 \pm 1$ and  $104 \pm 10 \pm 5$
$\pjpsi$ signals from $\EE$ annihilation in the $\yones$ and $\ytwos$ data samples, respectively.

We use the $N^{\pjpsi}_{\rm cont}$ obtained from the continuum data sample to calculate the cross section of the 
inclusive $\pjpsi$ production in $\EE$ annihilation via 
\beq
\sigma(\EE \to \pjpsi + anything) = \frac{N^{\pjpsi}_{\rm cont}}{L_{\rm cont} \times \eff_{\rm cont}^{\rm noP_c} 
\times \BR(\jpsi\to \LL) \times (1+\delta_{\rm ISR})}.
\eeq
Here, $\eff_{\rm cont}^{\rm noP_c} = 36.6\%$ is the efficiency obtained from no-$\pc$ MC simulation of continuum
production, and $\BR(\jpsi\to \LL) = (11.93 \pm 0.07)\%$ is the branching fraction of $\jpsi$ decaying to $\EE$ or
$\MM$~\cite{PDG}. For the inclusive production of hadronic final state $\pjpsi$ in the $\EE$ annihilation, we assume
a cross section $\propto 1/s$, taking reference to a measurement by CLEO on the total hadronic cross section in $\EE$
annihilation from $7.0~\gev$ to $10.5~\gev$~\cite{CLEO-R}. The radiative correction factor $(1+\delta_{\rm ISR})$ is
determined by $\int \sigma(s(1-x))F(x, s)dx/\sigma(s)$ and has the value 0.82, where $F(x, s)$ is the radiative
function obtained from QED calculation~\cite{rad01,isr02}. We obtain the cross section $\sigma(\EE \to \pjpsi +
anything) = (108\pm 11 \pm 6)~\fb$ at $\sqrt{s} = 10.52~\gev$, where the systematic uncertainties are discussed
in Sec.~\ref{syserr}.

\begin{figure}[tbp]
\centering
\psfig{file=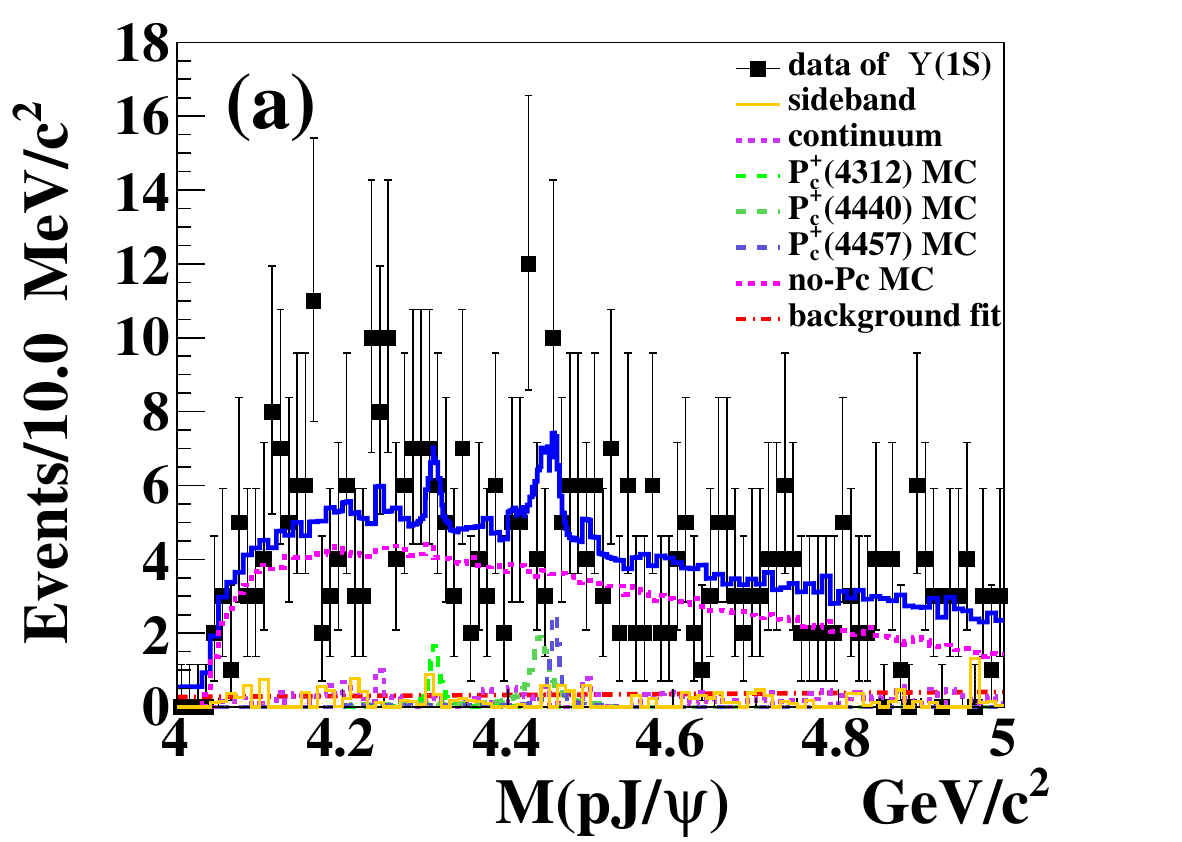,width=.75\textwidth}
\psfig{file=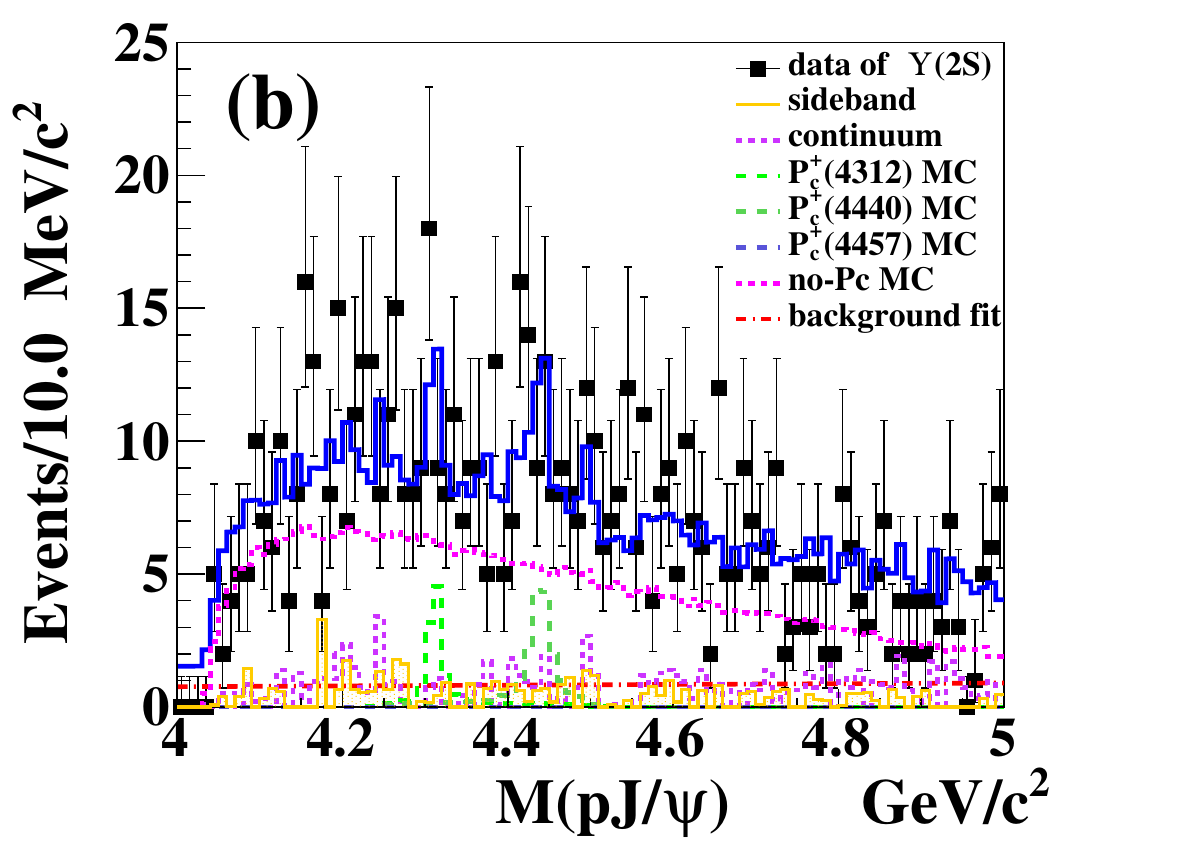, width=.75\textwidth}
\caption{The combined distributions of the invariant masses of $\pjpsi$ and $\pbjpsi$ from (a) the $\yones$
inclusive decays and (b) the $\ytwos$ inclusive decays, and the fit results including $\pca$, $\pcb$ and $\pcc$.
The dots with error bars are data. The brown solid histograms are the backgrounds estimated from the $\jpsi$
mass sidebands. The best fit results are shown in the blue solid curves, with different components being shown in
the dashed histograms and curves with different colors.}
\label{mpjpsi-sum}
\end{figure}

Figure~\ref{mpjpsi-sum} shows the combined distributions of Figs.~\ref{mpjpsi-bin2}(a-d) and \ref{mpjpsi-bin2}(e-h) 
for $\yones$ and $\ytwos$ inclusive decays, respectively.  We estimate the number of backgrounds from $K^\pm\jpsi$
or $\pi^\pm\jpsi$ to be $17.9\pm 1.2$ [$43.9\pm 3.0$] in $\yones$ [$\ytwos$] decays. With the backgrounds estimated
from the $\jpsi$ mass sidebands, mis-identification of proton, and continuum production being subtracted, we get
the final numbers of $\pjpsi$ signal events to be $N^{\pjpsi}_{\yones} = 377 \pm 24$ in the $\yones$ decays and
$N^{\pjpsi}_{\ytwos} = 564 \pm 32$ in the $\ytwos$ decays. These yields are much higher than those estimated due to
the underlying $\EE$ continuum production. To measure the production of $\pjpsi$ in $\yonetwos$ inclusive decays,
we use the no-$\pc$ MC samples to estimate the efficiencies to be $\eff_{\yonetwos}^{\rm noP_c} = 35.7\%$ and
36.2\% for the $\yones$ and $\ytwos$ inclusive decays. We calculate the branching fractions of $\yonetwos$
inclusive decays using
\beq
\BR[\yonetwos\to \pjpsi + anything] = \frac{N^{\pjpsi}_{\yonetwos} - f_{\rm scale}\times N^{\pjpsi}_{\rm cont}} 
{N_{\yonetwos} \times \eff_{\yonetwos}^{\rm noPc} \times \BR(\jpsi\to \LL)},
\eeq
where $N_{\yonetwos}$ are the numbers of $\yonetwos$ events in the $\yonetwos$ data samples. We obtain that 
$\BR[\yones\to \pjpsi + anything] = (8.1 \pm 0.6 \pm 0.5) \times 10^{-5}$ and $\BR[\ytwos\to
\pjpsi + anything] = (6.7 \pm 0.5 \pm 0.4) \times 10^{-5}$ for the first time. Taking into
account the branching fractions of the transitions from $\ytwos$ to $\yones$~\cite{PDG}, the $\BR[\yones\to \pjpsi
+ anything]$ measured here contributes a sub-branching fraction in the $\ytwos$ decays. With this sub-branching
fraction being subtracted, we obtain $\BR[\ytwos\to \pjpsi + anything] = (4.3 \pm 0.5 \pm 0.4)
\times 10^{-5}$.  Systematic uncertainties are listed in Table~\ref{tab-sys}, which is described in
Sec.~\ref{syserr}. The world average values of the branching fractions of $\jpsi$ production in $\yonetwos$ decays
are $\BR[\yones \to \jpsi + anything] = (5.4 \pm 0.4) \times 10^{-4}$ and $\BR[\ytwos\to \jpsi + anything] < 6
\times 10^{-3}$ at 90\% credibility~\cite{PDG}. Thus, the ratio $\BR(\Upsilon \to \pjpsi + anything)/\BR(\Upsilon
\to \jpsi + anything)$ is of order $10^{-2}-10^{-1}$ in $\yonetwos$ decays.

To estimate the production of a possible $\pc$ state in the $\yones$ or $\ytwos$ inclusive decays, we perform
unbinned maximum likelihood fits to the distribution of $M_{\pjpsi}$ in Fig.~\ref{mpjpsi-sum}(a) or
\ref{mpjpsi-sum}(b) with
\beq \label{pdf-fit1}
f_{\rm PDF} = f_{\pca} + f_{\pcb} + f_{\pcc} + f_{\rm noP_c} + f_{\rm cont}+ f_{\rm bkg},
\eeq
where $f_{\pca}$, $f_{\pcb}$, $f_{\pcc}$, $f_{\rm noP_c}$, and $f_{\rm cont}$ are the histogram PDFs obtained from
the signal MC simulations on $\pca$, $\pcb$, $\pcc$, the no-$\pc$ MC simulation, and the continuum data sample.
The component of $f_{\rm cont}$ is fixed according to the scale factor $f_{\rm scale}$.  We use a second-order
polynomial function for the $f_{\rm bkg}$ to describe the backgrounds due to $\jpsi$ selection. We fit the events
from the $\jpsi$ signal region with $f_{\rm PDF}$ and the events from $\jpsi$ mass sidebands with $f_{\rm bkg}$
simultaneously. The fit yields the numbers of $\pc$ signals [$N_{\rm fit}^{\rm A}(\pc)$], as listed in
Table~\ref{fit}. Since none of the $\pca$, $\pcb$, or $\pcc$ is significant, we integrate the likelihood versus
the $N_{\rm fit}^{\rm A}(\pc)$ and determine the upper limits $N_{\rm fit}^{\rm A,UL}(\pc)$ at 90\% credibility.
We also perform a fit to the $M_{\pjpsi}$ distribution in Fig.~\ref{mpjpsi-sum}(a) or ~\ref{mpjpsi-sum}(b) with
individual $\pc$ state in the $f_{\rm PDF}$, which yields the new number of $\pc$ signal
[$N_{\rm fit}^{\rm B}(\pc)$]. Similarly, we determine the related upper limits $N_{\rm fit}^{\rm B,UL}(\pc)$ for
$\pca$, $\pcb$, and $\pcc$ at 90\% credibility. We also estimate the upper limits by varying the masses and widths
of $\pc$ states by $1\sigma$ in these tests. Here, we consider the symmetric statistical uncertainties and the
asymmetric systematic uncertainties from the LHCb measurement~\cite{lhcb_pc_2019}. We take the largest values of
the upper limits as the conservative estimations of the upper limits of the numbers of the $\pc$ signals
[$N^{\rm UL}_{\rm sig}(\pc)$] in $\yonetwos$ inclusive decays. We then calculate the upper limit of the
branching fraction of a $\pc$ state produced in $\yones$ [$\ytwos$] inclusive decays at 90\% credibility with 
\beq\label{br-pc}
\BR[\yonetwos \to \pc + anything] \cdot \BR(\pc\to\pjpsi) < \frac{N^{\rm UL}_{\rm sig}(\pc)}{N_{\yonetwos}\cdot 
\eff^{\rm MC}_{\rm allcmb} \cdot \BR(\jpsi\to \LL)(1-\delta_{\rm sys})},
\eeq
where $\delta_{\rm sys} = 6.1\%$ (5.9\%) is the systematic uncertainty of $\yones$ [$\ytwos$] decays, which are
described in Sec.~\ref{syserr}. We summarize the values of $N_{\rm fit}^{\rm A}(\pc)$,
$N_{\rm fit}^{\rm A,UL}(\pc)$, $N_{\rm fit}^B(\pc)$, $N_{\rm fit}^{\rm B,UL}(\pc)$, $N^{\rm UL}_{\rm sig}(\pc)$,
and the upper limit of $\BR[\yonetwos \to \pc + anything] \cdot \BR(\pc\to\pjpsi)$ at 90\% credibility in
Table~\ref{fit}.

\begin{table}
\begin{center}
\caption{The fit results and the upper limits of $\pca$, $\pcb$, and $\pcc$ productions in $\yonetwos$ inclusive 
decays. $N_{\rm fit}^{\rm A}$ is the number of $\pc$ signals in the fit with the PDF function $f_{\rm PDF}$ contains 
$\pca$, $\pcb$, and $\pcc$ states, and $N_{\rm fit}^{\rm A,UL}$ is the related upper limits at 90\% credibility. 
$N_{\rm fit}^{\rm B}$ is the number of $\pc$ signals in the fit with the PDF function that contains only a single 
$\pc$ state, and $N_{\rm fit}^{\rm B,UL}$ is the related upper limits at 90\% credibility. $N^{\rm UL}_{\rm sig}$ 
is the final conservative estimation of the upper limit of the number of $\pc$ signals in $\yonetwos$ inclusive decays. 
$\BR^{\rm UL}$ is the upper limit of $\BR(\Upsilon\to \pc + anything)\cdot \BR(\pc \to \pjpsi)$ at 90\%
credibility.}
\begin{small}
\begin{tabular}{c | c | c | c | c | c | c }\hline\hline 
        					&\multicolumn{3}{c|}{$\yones$ decays} &  \multicolumn{3}{c}{$\ytwos$ decays} \\\hline
   --- 						& $\pca$  & $\pcb$ & $\pcc$ & $\pca$ & $\pcb$ & $\pcc$   \\\hline
$N_{\rm fit}^{\rm A}$ &  $4 \pm 8$   & $10\pm 11$  & $7 \pm 9$  & $19\pm 14$ & $30\pm 16$ & $2\pm 11$       \\
$N_{\rm fit}^{\rm A,UL}$    &  18   & 28   & 22  & 43  & 58  & 15       \\
$N_{\rm fit}^{\rm B}$ &  $8 \pm 7$   & $10 \pm 10$  & $7 \pm 8$  & $20\pm 11$  & $26 \pm 12$ & $3\pm 11$      \\
$N_{\rm fit}^{\rm B,UL}$  &  22   & 26  & 27 & 42  & 50 & 13        \\
$N^{\rm UL}_{\rm sig}$  &  26   & 45  & 34 & 51  & 83 & 32       \\\hline
$\BR^{\rm UL}$ ($\times 10^{-6}$) & 
  5.7 & 10.0 & 7.6 & 7.2 & 11.8 & 4.6 \\\hline\hline
\end{tabular}
\end{small}
\label{fit}
\end{center}
\end{table}

According to the distributions shown in Fig.~\ref{mpjpsi-bin2} (i-l), the continuum production of $\pjpsi$ is low
and the background level due to $\jpsi$ selection is high. We check the sum of these $M_{\pjpsi}$ distributions
after the backgrounds have been subtracted and see no obvious $\pc$ signal. Therefore, we do not search for a
pentaquark state in the continuum data sample.

\section{Systematic uncertainties}
\label{syserr}

\begin{table}
\caption{The summary of the systematic uncertainties (\%)  in the measurements of $\BR[\yonetwos\to \pjpsi + 
anything]$ and $\sigma(\EE \to \pjpsi + anything)$ at $\sqrt{s} = 10.52~\gev$.}
\label{tab-sys}
\begin{center}
\begin{tabular}{c | c | c | c}
\hline\hline 
Source     & $\yones$ decay & $\ytwos$ decay & $\sigma(\EE \to \pjpsi + anything)$\\
\hline
Particle identification                       &  2.1 & 2.1 & 2.1\\
Tracking                  &  1.1 & 1.1 & 1.1\\
$\jpsi$ signal region     &  0.5 & 0.4 & 0.2 \\
$\MMS(\pjpsi)$ requirement & 0.4 & 0.9 & 2.1 \\
$\BR(\jpsi\to \LL)$       &  0.6 & 0.6 & 0.6\\
$1+\delta_{\rm ISR}$ & --- & --- & 1.0 \\
Scale factor $f_{\rm scale} $  & 0.8 & 1.1 & --- \\
Modeling in MC simulation & 4.6 & 4.2 & 4.5 \\
Number of $\yonetwos$ events &  2.2 & 2.3 & ---\\
Integrated luminosity    &  --- & --- & 1.4\\
Statistics of MC samples  &  0.5 & 0.5 & 0.5\\
\hline
Sum in quadrature         & 5.8  & 5.6 & 5.8 \\
\hline \hline 
\end{tabular}
\end{center}
\end{table}

As listed in Table~\ref{tab-sys}, we consider the following systematic uncertainties in determining the branching 
fractions $\BR[\yonetwos\to \pjpsi + anything]$ and measuring $\sigma(\EE \to \pjpsi + anything)$ at $\sqrt{s} = 
10.52~\gev$: particle identification and mis-identification, tracking efficiency, $\jpsi$ signal region,
$\MMS(\pjpsi)$ requirement, branching fraction of $\jpsi$ decay, radiative correction factor, scale factor
$f_{\rm scale}$, modeling in MC simulation, number of $\yonetwos$ events, integrated luminosity, and statistics of
MC samples, etc. The uncertainties due to the lepton identification are 2.0\% and 0.5\% for $e^\pm$ and $\mu^\pm$,
respectively. For proton identification, we applied an efficiency correction according to the momentum and angle in
the laboratory frame. Shifting the correction factor by $1\sigma$, we get the related efficiency difference of
0.43\% and take 0.5\% to be the systematic uncertainty of proton identification. Therefore, the total systematic
uncertainty due to the particle identification is 2.1\%. In estimating the backgrounds from $K\jpsi$ or $\pi\jpsi$,
the mis-identification of $K(\pi)$ to $p$ is ($1.98 \pm 0.07)\% ~ [(0.72 \pm 0.02)\%]$. The uncertainties of
mis-identification contribute 0.4, 1.1, 1.3 in the numbers of estimated backgrounds from $K\jpsi$ and $\pi\jpsi$ in
$\yones$ decays, $\ytwos$ decays, and continuum productions. They are tiny comparing to the number of $\pjpsi$
signal events and not listed in Table~\ref{tab-sys}. The uncertainty due to the tracking efficiency is 0.35\% per
track and adds linearly. Fitting the $M_{\LL}$ distributions from data and MC simulations with a Gaussian function
for $\jpsi$ signal and a second-order Chebychev function for backgrounds, we obtain the efficiencies of $\jpsi$ mass
signal window to be $\eff^{\rm data}_{\jpsi} = (99.43 \pm 0.58)\%$, $(99.56 \pm 0.48)\%$, and $(99.69 \pm 0.37)$\%
in $\yones$ decays, $\ytwos$ decays, and continuum productions in data, and $\eff^{\rm MC}_{\jpsi} = 99.9\%$ in
the signal MC simulations. We take the differences of the efficiencies in data and MC simulation to be the
systematic uncertainties, i.e., 0.5\% in the $\yones$ decays, 0.4\% in the $\ytwos$ decays, and 0.2\% in the
continuum productions. To estimate systematic uncertainties of the requirement $\MMS(\pjpsi) > 10~\gevcss$, we set
$\mrec^2(\jpsi) > 20~(\gevcs)^2$ to suppress the sideband background and then calculate efficiencies. The
efficiencies are 99.5\% (99.1\%), 98.3\% (99.2\%) and 97.7\% (99.8\%) for the data~(MC) of $\yones$, $\ytwos$ and
continuum, respectively. The systematic uncertainties of $\yones$, $\ytwos$ and continuum are respectively 0.4\%,
0.9\% and 2.1\%. According to the world average values~\cite{PDG}, $\BR(\jpsi\to \LL)(l=e,\mu)$ contributes a
systematic uncertainty of 0.6\%. The precision of calculating the factor $1+\delta_{\rm ISR}$ is
0.2\%~\cite{rad01, isr02}. Additionally, by varying the photon energy cutoff by $50~\mev$ in the simulation of ISR,
we determine the change of $1+\delta_{\rm ISR}$ to be 0.01 and take 1.0\% to be the conservative systematic
uncertainty in measuring the cross section $\sigma(\EE \to \pjpsi + anything)$ at $\sqrt{s} = 10.52~\gev$. By
changing $s_{\rm cont}/s_{\ytwos}$ to $[s_{\rm cont}/s_{\ytwos}]^{3/2}$, the value of $f_{\rm scale}$ changes from
0.077 to 0.086 in $\yones$ decays, and from 0.303 to 0.318 in $\ytwos$ decays. We take 11.2\% and 4.9\% as their
systematic uncertainties, but they are not considered in estimating the upper limits of $\pc$ productions. They
correspond to 0.8\% and 1.1\% in measuring the branching fractions of $\yones$ and $\ytwos$ decays, respectively.
There is no theoretical calculation for the inclusive production of $\pjpsi$ in $\yonetwos$ decays or $\EE$
annihilation, therefore, we choose the PHSP model in the MC simulations. We find a reasonable agreement between
data and MC simulations in the $\mrec^2(\pjpsi)$ distribution, in which we apply a selection. We also get good
agreements between data and MC simulations in many distributions, such as the $\mrec^2(\jpsi)$ and the proton
momentum. To estimate the uncertainties in modeling the final states in the MC simulations, we vary the mass and
width of $X$ by $200~\mevcs$ and $500~\mev$ in the hadronization of $\qqb$, which have differences in efficiency
that 2.3\% in the $\yones$ decays, 1.9\% in the $\ytwos$ decays and 2.2\% in the continuum production,
respectively. Considering that the proton candidate may come from $\Lambda$ decay, we simulate the MC samples of
$\yonetwos\to \pjpsi + \bar{\Lambda} + (s\bar{u})$ and find the efficiency differences, from those of $\pc$ signal
MC samples, of 3.8\% in $\yones$ decays, 3.6\% in $\ytwos$ decays, and 3.8\% in continuum production. The various combination 
ratios derived from the data are adjusted by $1\sigma$ of their statistical uncertainties to modify the relative fractions of the 
$\jpsi p\pbar$ and $\jpsi p \bar{n}$  channels within the mixed MC samples. The resulting changes in efficiencies are considered 
systematic uncertainties when combining the MC samples. These systematic uncertainties are 1.1\% for $\yones$ decays, 
0.8\% for $\ytwos$ decays, and 0.7\% for continuum production. We sum the
these sources and obtain the systematic uncertainties in modeling the final states in MC simulations to be 4.6\%,
4.2\%, and 4.5\% in $\yones$ decays, $\ytwos$ decays, and continuum productions at $\sqrt{s} = 10.52~\gev$. The
uncertainties of the total numbers of $\yones$ events and $\ytwos$ events are 2.2\% and 2.3\% in the Belle data
samples~\cite{y1sevnt, y2sevnt}. The common uncertainty in the integrated luminosities for the $\yones$, $\ytwos$,
and continuum data samples is 1.4\%, which is canceled in calculating the scale factor $f_{\rm scale}$. The
statistical uncertainties of the signal MC samples are 0.5\% in common. Assuming these uncertainties are
independent and sum them in quadrature, we obtain the total systematic uncertainties to be 5.8\% in $\BR[\yones\to
\pjpsi + anything]$, 5.6\% in $\BR[\ytwos\to \pjpsi + anything]$, and 5.8\% in $\sigma(\EE \to \pjpsi + anything)$
at $\sqrt{s} = 10.52~\gev$. 

In determining the upper limits of $\pc$ productions in $\yonetwos$ decays, most of the systematic uncertainties are 
the same as those listed in Table~\ref{tab-sys}, except for the modeling of $\pjpsi$ in signal MC simulations and
additional uncertainties in fits. To evaluate these, we do similar studies, including varying the mass and width of
$X\to \qqb$ and simulating the MC sample of $\yonetwos\to \pc + \bar{\Lambda} + (s\bar{u})$. We replace the
uncertainties in modeling by 5.1\% in $\yones$ decays and 4.7\% in $\ytwos$ decays and eliminate those of
$f_{\rm scale}$ in Table~\ref{tab-sys}. Therefore, the total systematic uncertainties of $\pc$ productions in
$\yones$ decays and $\ytwos$ decays are $\delta_{\rm sys} = 6.1\%$ and 5.9\%, respectively. To estimate the
systematic uncertainty of $f_{\rm noP_c}$ in the fits, we investigate the difference in the yield when using an
ARGUS function to replace the histogram PDF obtained from the no-$\pc$ MC simulation~\cite{argusfunction}. We
change the masses and the widths of the $\pc$ states by $1\sigma$ according to LHCb
measurement~\cite{lhcb_pc_2019}. In calculating the value of $\sigma$, we take into account the symmetric
statistical uncertainty and the asymmetric systematic uncertainties. To estimate the systematic uncertainty of
$f_{\rm cont}$ in the fit, we vary the factor $f_{\rm scale}$ by $1\sigma$. As before, we take the highest
values of $N^{\rm UL}_{\rm sig}(\pc)$ to calculate the upper limit of $\pc$ production in the $\yonetwos$ inclusive
decays.

\section{Summary}

We study the $\pjpsi$ final states in $\yonetwos$ inclusive decays and search for the $\pca$, $\pcb$ and $\pcc$ 
signals. To study the production of $\pjpsi$ in the $\yonetwos$ data samples, we also investigate the $\pjpsi$ final 
state in the Belle continuum data sample. We determine the branching fractions to be $\BR[\yones\to \pjpsi +
anything] = (8.1\pm 0.6 \pm 0.5) \times 10^{-5}$ and $\BR[\ytwos \to \pjpsi + anything] = (4.3 \pm 0.5 \pm 0.4) \times 10^{-5}$, and the cross section of continuum production to be $\sigma(\EE \to \pjpsi + anything) =
(108 \pm 11 \pm 6)~\fb$ at $\sqrt{s} = 10.52~\gev$. Here, the branching fractions $\BR[\ytwos \to \pjpsi + anything]$ exclude the decays via $\yones$. No significant $\pc$ signals exist in the Belle
$\yonetwos$ data samples. We determine the upper limits of $\pc$ productions in $\yonetwos$ inclusive decays to be 
\beqar 
\BR[\yones\to \pca + anything]\cdot \BR[\pca\to \pjpsi] & < & 5.7 \times 10^{-6}, \\
\BR[\yones\to \pcb + anything]\cdot \BR[\pcb\to \pjpsi] & < & 10.0 \times 10^{-6}, \\
\BR[\yones\to \pcc + anything]\cdot \BR[\pcc\to \pjpsi] & < & 7.6 \times 10^{-6}, \\
\BR[\ytwos\to \pca + anything]\cdot \BR[\pca\to \pjpsi] & < & 7.2 \times 10^{-6}, \\
\BR[\ytwos\to \pcb + anything]\cdot \BR[\pcb\to \pjpsi] & < & 11.8 \times 10^{-6}, \\
\BR[\ytwos\to \pcc + anything]\cdot \BR[\pcc\to \pjpsi] & < & 4.6 \times 10^{-6},
\eeqar
at 90\% credibility.

\acknowledgments

This work, based on data collected using the Belle detector, which was
operated until June 2010, was supported by 
the Ministry of Education, Culture, Sports, Science, and
Technology (MEXT) of Japan, the Japan Society for the 
Promotion of Science (JSPS), and the Tau-Lepton Physics 
Research Center of Nagoya University; 
the Australian Research Council including grants
DP210101900, 
DP210102831, 
DE220100462, 
LE210100098, 
LE230100085; 
Austrian Federal Ministry of Education, Science and Research (FWF) and
FWF Austrian Science Fund No.~P~31361-N36;
National Key R\&D Program of China under Contract No.~2022YFA1601903,
National Natural Science Foundation of China and research grants
No.~11575017,
No.~11761141009, 
No.~11705209, 
No.~11975076, 
No.~12135005, 
No.~12150004, 
No.~12161141008, 
and
No.~12175041, 
and Shandong Provincial Natural Science Foundation Project ZR2022JQ02;
the Czech Science Foundation Grant No. 22-18469S;
Horizon 2020 ERC Advanced Grant No.~884719 and ERC Starting Grant No.~947006 ``InterLeptons'' (European Union);
the Carl Zeiss Foundation, the Deutsche Forschungsgemeinschaft, the
Excellence Cluster Universe, and the VolkswagenStiftung;
the Department of Atomic Energy (Project Identification No. RTI 4002), the Department of Science and Technology of India,
and the UPES (India) SEED finding programs Nos. UPES/R\&D-SEED-INFRA/17052023/01 and UPES/R\&D-SOE/20062022/06; 
the Istituto Nazionale di Fisica Nucleare of Italy; 
National Research Foundation (NRF) of Korea Grant
Nos.~2016R1\-D1A1B\-02012900, 2018R1\-A2B\-3003643,
2018R1\-A6A1A\-06024970, RS\-2022\-00197659,
2019R1\-I1A3A\-01058933, 2021R1\-A6A1A\-03043957,
2021R1\-F1A\-1060423, 2021R1\-F1A\-1064008, 2022R1\-A2C\-1003993;
Radiation Science Research Institute, Foreign Large-size Research Facility Application Supporting project, the Global Science Experimental Data Hub Center of the Korea Institute of Science and Technology Information and KREONET/GLORIAD;
the Polish Ministry of Science and Higher Education and 
the National Science Center;
the Ministry of Science and Higher Education of the Russian Federation
and the HSE University Basic Research Program, Moscow; 
University of Tabuk research grants
S-1440-0321, S-0256-1438, and S-0280-1439 (Saudi Arabia);
the Slovenian Research Agency Grant Nos. J1-9124 and P1-0135;
Ikerbasque, Basque Foundation for Science, and the State Agency for Research
of the Spanish Ministry of Science and Innovation through Grant No. PID2022-136510NB-C33 (Spain);
the Swiss National Science Foundation; 
the Ministry of Education and the National Science and Technology Council of Taiwan;
and the United States Department of Energy and the National Science Foundation.
These acknowledgements are not to be interpreted as an endorsement of any
statement made by any of our institutes, funding agencies, governments, or
their representatives.
We thank the KEKB group for the excellent operation of the
accelerator; the KEK cryogenics group for the efficient
operation of the solenoid; and the KEK computer group and the Pacific Northwest National
Laboratory (PNNL) Environmental Molecular Sciences Laboratory (EMSL)
computing group for strong computing support; and the National
Institute of Informatics, and Science Information NETwork 6 (SINET6) for
valuable network support.

\end{document}